\documentclass[aps,pra,preprint,superscriptaddress]{revtex4-1}  
\usepackage{array}
\usepackage{booktabs}
\usepackage{siunitx}
\usepackage{amsfonts}
\usepackage{amsmath}
\usepackage{amssymb}
\usepackage{graphicx}
\usepackage{mathrsfs} 
\usepackage{color} 
\usepackage{floatflt}

\providecommand\bx{\mathbf{x}}

\providecommand\be{\mathbf{\hat{e}}}
\providecommand\bt{\mathbf{\hat{t}}}
\providecommand\bn{\mathbf{\hat{n}}}

\newcommand{\pd}[2]{\frac{\partial #1}{\partial #2}}

\newcommand{\ub}[1]{^{({#1})}}

\begin{document}
\title{Geometric quantization of localized surface plasmons}
\author{Ory Schnitzer}
\affiliation{Department of Mathematics, Imperial College London, 180 Queen's Gate, SW7 2AZ, London, UK}
\begin{abstract}
We consider the quasi-static problem governing the localized surface plasmon modes and permittivity eigenvalues $\epsilon$ of smooth, arbitrarily shaped, axisymmetric inclusions. We develop an asymptotic theory for the dense part of the spectrum, i.e., close to the accumulation value $\epsilon=-1$ at which a flat interface supports surface plasmons; in this regime, the field oscillates rapidly along the surface and decays exponentially away from it on a comparable scale. With $\tau=-(\epsilon+1)$ as the small parameter, we develop a  surface-ray description of the eigenfunctions in a narrow boundary layer about the interface; the fast phase variation, as well as the slowly varying amplitude and geometric phase, along the rays are determined as functions of the local geometry. We focus on modes varying at most moderately in the azimuthal direction, in which case the surface rays are meridian arcs that focus at the two poles. Asymptotically matching the diverging ray solutions with expansions valid in inner regions in the vicinities of the poles yields the quantization rule
$$\frac{1}{\tau} \sim \frac{\pi n }{\Theta}+\frac{1}{2}\left(\frac{\pi}{\Theta}-1\right)+o(1),$$ 
where $n\gg1$ is an integer and $\Theta$  a geometric parameter given by the product of the inclusion length and the reciprocal average of its cross-sectional radius along its symmetry axis. For a sphere, $\Theta=\pi$, whereby the formula returns the exact eigenvalues $\epsilon_n=-1-1/n$. We also demonstrate good agreement with exact solutions in the case of prolate spheroids. 
\end{abstract}

\maketitle

\section{Introduction}\label{sec:intro}
In nanoplasmonics, the surface-plasmon resonances of metallic nanoparticles and nanostructures are exploited in order to manipulate light on nanometric, subwavelength, scales \cite{Maier:07,Schuller:10,Luk:10}.  The plasmonic eigenvalue problem is fundamental to this area of applied physics, as it determines the plasmonic eigenmodes supported by a given inclusion, or cluster of inclusions \cite{Ouyang:89,Fredkin:03,Mayergoyz:05,Klimov:14,Davis:17}. The same spectral problem arises in several other areas, such as in the theory of composite media \cite{Bergman:79}; the plasmonic spectrum is also closely linked to the specturm of the Neumann--Poincar{\'e} integral operator \cite{Grieser:14,Ando:16,Ammari:17}. There is an enormous body of literature on this problem, with recent developments including analyses of strongly interacting particles using separation of variables \cite{Klimov:07,Lebedev:13}, transformation optics \cite{Pendry:13}, multipole methods \cite{Yu:18}, matched asymptotic expansions \cite{Schnitzer:15plas,Schnitzer:18} and layer-potential techniques \cite{Ammari:17,Bonnetier:18}; analysis of corners \cite{Perfekt:14,Bonnetier:17}; application to stimulated emission \cite{Bergman:03} and second-harmonic generation \cite{Li:05}; regular shape perturbations \cite{Grieser:09,Ando:18}; extensions incorporating nonlocality \cite{Luo:13,Raza:15,Schnitzer:16,Schnitzer:16b} and retardation \cite{Bergman:80,Agranovich:99,Farhi:15,Ammari:16,Chen:17}; and high-mode-number asymptotics \cite{Ando:16exp,Miyanishi:18}. 

The plasmonic eigenvalue problem naturally arises when looking for nontrivial time-harmonic solutions of Maxwell's equations, in the absence of any external forcing and in the limit where the inclusion is small compared to the electromagnetic wavelength. In the latter regime, a quasi-static description holds in the vicinity of the inclusion, whereby the field can be derived from a potential $\varphi$. The plasmonic eigenvalue problem is then to find permittivity eigenvalues $\epsilon$, defined here relative to the background medium, and corresponding eigenfunctions $\varphi$ that satisfy Laplace's equation
\begin{equation}\label{laplace}
\nabla^2\varphi=0
\end{equation}
in the inclusion and background domains, together with the transmission conditions 
\begin{equation}\label{transmission}
\varphi\ub{i}=\varphi\ub{o}, \quad \epsilon\pd{\varphi\ub{i}}{n}=\pd{\varphi\ub{o}}{n}
\end{equation}
on the inclusion boundary, where superscripts $(i)$ and $(o)$ henceforth refer to the inside and outside of the inclusion, respectively. We also require that the potentials attenuate,
\begin{equation}\label{decay}
\varphi\to0,
\end{equation}
at large distances from the inclusion; while it's actually the gradient that must attenuate, so that the near field matches with outward propagating solutions of Maxwell's equations, the present formulation is convenient as it eliminates an immaterial additive freedom.

The plasmonic eigenvalues are scale invariant and depend solely on the geometry of the inclusion; for smooth geometries, the spectrum consists of an infinite discrete set of negative real eigenvalues accumulating at $\epsilon=-1$. The associated set of modes can be used to rigorously expand, for example in the electromagnetic context, the near field induced by arbitrary external forcing and for an arbitrary --- complex valued and frequency dependent --- inclusion permittivity \cite{Klimov:14,Davis:17}. In particular, for metallic inclusions in the visible regime, the real part of the inclusion permittivity is negative while the imaginary part, which accounts for ohmic losses, is often relatively small; at frequencies where the inclusion permittivity is near to one of the eigenvalues, the surface-plasmon mode which correspond to it may be resonantly excited, depending on the level of losses and overlap between the external forcing and that mode. The subwavelength nature of the surface-plasmon modes is also crucial in this context, as radiation losses are negligible in this regime \cite{Maier:07}.  

It is usually the lowest-order plasmonic modes that are resonantly excited; for example, only the three ``dipolar'' modes of a subwavelength sphere, with eigenvalue $\epsilon=-2$, are excited by an incident plane wave \cite{Maier:07}. In the other extreme, there are the high-mode-number modes with eigenvalues near the accumulation point of the spectrum. As the accumulation value is approached the eigenfunctions oscillate along the boundary, and decay away from it, on shorter and shorter length scales. Such high-mode-number (or large-wavenumber) modes can only be excited by external sources in the near field, close to the inclusion boundary; for realistically lossy inclusions, they are typically not manifested as isolated resonances, given the high density of states, but rather contribute collectively. Theoretically, they are known to play a dominant role in some of the most fascinating effects discussed in relation to plasmonics, such as perfect lensing \cite{Pendry:00,Larkin:05,Farhi:14}, anomalous localized resonance \cite{Milton:06} and Van der Waals interactions \cite{Klimov:09,luo:14}. 

In three dimensions, the convergence of the eigenvalues to the accumulation value is usually algebraic; e.g., for a sphere geometry the unique eigenvalues are exactly 
\begin{equation}\label{ev sphere}
\epsilon=-1-\frac{1}{n},
\end{equation} 
where $n\in\mathbb{N}$. Recently, Miyanishi \cite{Miyanishi:18} proved a Weyl's law for arbitrarily shaped smooth inclusions in three dimensions, providing the unsigned leading-order asymptotics of the ordered eigenvalues counting multiplicities in terms of two global geometric properties: the Euler characteristic and Willmore energy.  This remarkable mathematical result unfortunately does not provide any connection between the asymptotic eigenvalues and their corresponding eigenfunctions, nor asymptotic approximations for those  eigenfunctions. Furthermore, this leading order result it is not accurate enough to distinguish between adjacent eigenvalues, nor does it provide the asymptotics for the set of unique eigenvalues without prior knowledge of the degeneracy (e.g., to compare with \eqref{ev sphere} it is necessary to know that for spheres the $n$th unique eigenvalue is $2n+1$ degenerate). In the plasmonics context, asymptotic solutions of the eigenvalue problem for high mode numbers could potentially be used to study the optical excitation of a nano-metallic particle in scenarios where the collective response of the high-order modes dominates \cite{Milton:06,Farhi:14,Farhi:17}. To this end, it is necessary to find at least leading-order asymptotic approximations for the eigenfunctions, as well as accurate eigenvalue asymptotics, up to an order depending on the smallness of the imaginary part of the inclusion permittivity. 

In this paper, we put forward a novel approach for finding asymptotic solutions of the plasmonic eigenvalue problem in the high-mode-number limit. Our approach is based on the observation that high-order modes are strongly confined to the boundary and oscillate rapidly along it; they are locally tantamount to surface plasmons of a flat interface. We propose that singular perturbation techniques \cite{Hinch:91}, namely matched asymptotics and WKBJ-type expansions, could be used to develop a surface-plasmon-ray description for the eigenfunctions of smooth, arbitrarily shaped, inclusions, together with quantization rules governing the corresponding eigenvalues. Our approach is inspired by the geometric theory of diffraction \cite{Keller:62} and its application to finding  asymptotic solutions to eigenvalue problems \cite{Keller:58,Keller:60}. To demonstrate this approach, we focus herein on one family of shapes, smooth axisymmetric inclusions, that is general yet relatively easily addressed. For simplicity, we assume that the thickness is a single-valued function of the axial co-ordinate, that the tips are locally spheroidal; furthermore, among all the large-wavenumber modes we focus on those that vary moderately in the azimuthal direction. We were led to this work after studying the exact eigenfunctions of spheres and realizing their relation to surface plasmons of a flat interface. We accordingly find it useful to preface our asymptotic analysis with a brief discussion of this linkage.

\textbf{Flat surfaces and spheres.}
\begin{figure}[t!]
\begin{center}
\includegraphics[scale=0.45]{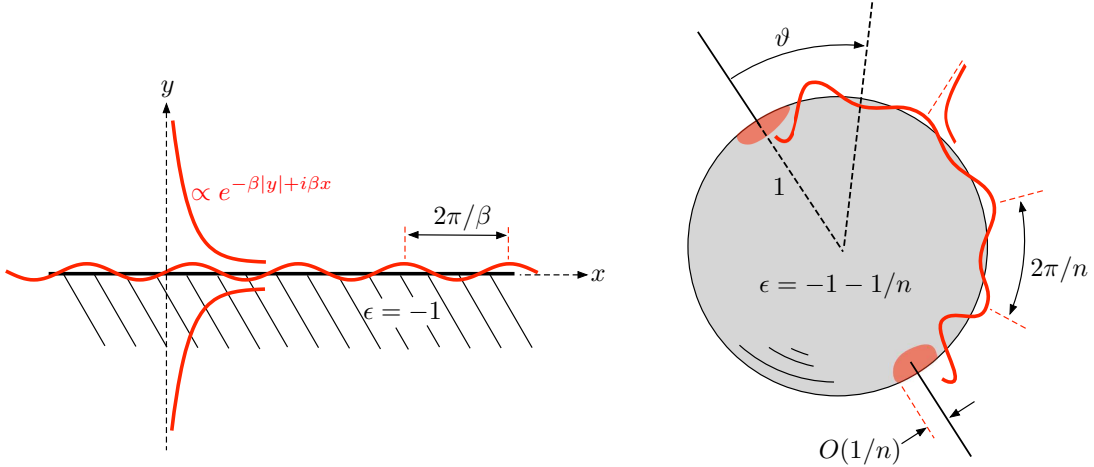}
\caption{(a) Surface plasmons at a flat interface. (b) A high-mode number (or large-wavenumber) axisymmetric localized surface plasmon mode of a sphere.}
\label{fig:plasmon}
\end{center}
\end{figure}
Consider first an infinite plane interface, say $y=0$, separating two homogeneous half spaces. Looking for surface-wave solutions of \eqref{laplace}--\eqref{decay} in the form $\varphi=\psi(y)\exp(i\beta x)$, we find $\epsilon\equiv-1$ with eigenfunctions $\psi(y)\propto\exp(-\beta |y|)$ (see figure \ref{fig:plasmon}(a)); here $\beta$ is a real wavenumber and $(x,y)$ are Cartesian co-ordinates along and normal to the interface, respectively. Physically, these solutions represent  surface plasmons, i.e., quasi-static collective oscillations of surface-polarization charge and electric field; surface plasmons are the limiting case of electromagnetic surface waves (surface plasmon polaritons) as the plasmon wavelength, $2\pi/\beta$, becomes small compared to the wavelength of light in the bulk media \cite{Maier:07}. Turning to the case of smooth inclusions, it becomes clear that modes that oscillate rapidly along the boundary of the particle (and hence are strongly confined to it) must correspond to  eigenvalues near the accumulation value, since on the small scale of the oscillations the smooth surface is approximately flat. 

With the above surface-plasmon solutions in mind, let us now inspect the modes of a sphere, which are known in closed form \cite{Klimov:14}.  The exact expression \eqref{ev sphere} immediately reveals the high-mode-number asymptotics of the eigenvalues. To describe the eigenfunctions, we scale the sphere radius to unity and introduce spherical co-ordinates $(r,\vartheta,\phi)$, where for each $n$ there are modes having azimuthal dependence $\exp(im\phi)$, where $m=-n,\ldots,n$. In the axisymmetric case, $m=0$, the eigenfunctions are $\varphi = r^{-(n+1)}P_{n}(\cos\vartheta)$ for $r>1$ and $r^n P_{n}(\cos\vartheta)$ for $r<1$, where $P_n$ are the Legendre polynomials. 
In the high-mode number limit, the boundary localization transitions from being algebraic to exponential: 
\begin{equation}
\varphi \sim e^{-n|r-1|}P_n(\cos\vartheta) \quad \text{as} \quad n\to\infty.
\end{equation}
The dependence of the axisymmetric eigenfunctions on the polar angle $\vartheta$ is more subtle. In the ``outer'' limit, namely with $\vartheta$ fixed, 
\begin{equation}\label{sphere outer}
P_n(\cos\vartheta)\sim \left(\frac{2}{\pi n \sin\vartheta}\right)^{1/2}\cos\left[(n+1/2)\vartheta-\pi/4\right] \quad \text{as} \quad n\to\infty.
\end{equation}
As expected, the mode is locally a surface plasmon of a flat interface, with a constant wavenumber $\beta\approx n$; note the slow amplitude variation on the scale of the radius, as well as the comparably slow phase accumulation, corresponding to an $O(1)$ correction to the wavenumber. Clearly, the outer approximation \eqref{sphere outer} diverges at the two poles; there is an  ``inner region'' near to the pole at $\vartheta=0$, 
\begin{equation}
\vartheta=O(1/n): \quad P_n(\cos\vartheta)\sim J_0(n\vartheta)  \quad \text{as} \quad n\to\infty,
\end{equation}
with a similar expansion holding near $\vartheta=\pi$ (see figure \ref{fig:plasmon}(b)). Note the relative enhancement of the eigenfunctions at the poles. 

\section{Smooth axisymmetric inclusions}\label{sec:formulation}
\subsection{Geometry}\label{ssec:geometry}
\begin{figure}[t!]
\begin{center}
\includegraphics[scale=0.7]{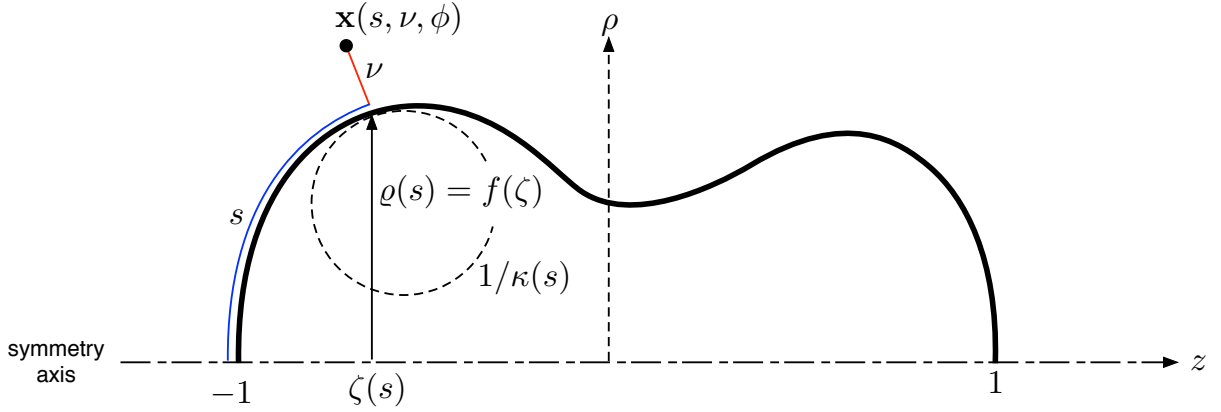}
\caption{Dimensionless schematic of the axisymmetric geometry.}
\label{fig:schematic1}
\end{center}
\end{figure}

Consider a smooth axisymmetric inclusion of length $2a$; for simplicity, we assume that the inclusion's thickness is a single-valued function of the axial co-ordinate and that the tips are locally spheroidal. In light of the scale invariance of the plasmonic eigenvalue problem, it is natural to adopt a dimensionless convention where lengths are normalised by $a$. Using  cylindrical co-ordinates $(\rho,\phi,z)$, 
the inclusion boundary can be defined as 
\begin{equation}
\rho=f(z), \quad -1 \le z \le1,
\end{equation}
where $f(z)$ is a smooth thickness profile that is positive for $-1<z<1$ and vanishes at the tips $z=\pm1$; the assumption that the tips are locally spheroidal implies 
\begin{equation}\label{tips}
f^2\sim \frac{2}{\kappa_l}(z+1) \quad \text{as} \quad z\to-1; \quad f^2\sim \frac{2}{\kappa_r}(1-z) \quad \text{as} \quad z\to1,
\end{equation} 
where $\kappa_l,\kappa_r>0$ are shape constants encapsulated in $f(z)$. Figure \ref{fig:schematic1} shows a dimensionless schematic of the geometry.

\subsection{Symmetry and boundary-fitted co-ordinates}\label{ssec:symmetry}
Since the eigen-potentials $\varphi$ are defined up to an arbitrary multiplicative factor, we may consider them to be dimensionless. As discussed in the introduction, the plasmonic eigenvalue problem consists of Laplace's equation \eqref{laplace}, the transmission conditions \eqref{transmission}, as well as attenuation at large distances. For  axisymmetric inclusions, symmetry implies that the eigen-potentials can be sought in the form
\begin{equation}
\varphi=\psi e^{im\phi}, \quad m\in \mathbb{Z},
\end{equation}
where $\psi$ is independent upon the azimuthal angle $\phi$. It will be convenient to use orthogonal boundary-fitted coordinates $(s,\nu,\phi)$, which can be defined in some finite neighbourhood of the boundary. On the boundary, the normal coordinate $\nu$ vanishes and $s$ coincides with the arc length measured along a meridian arc starting from $z=-1$; we take $\nu$ to be positive in the domain external to the particle. As shown in figure \ref{fig:schematic1}, we denote $\rho=\varrho(s)$ and $z=\zeta(s)$ on $\nu=0$, and let $\kappa(s)$ be the signed curvature of a meridian arc, positive when the center of curvature is inside the inclusion. We show in the appendix that, in a neighbourhood of the boundary where the boundary-fitted co-ordinates are  defined, Laplace's equation \eqref{laplace} is
\begin{multline}\label{fitted laplace}
\pd{}{s}\left(\frac{\varrho+\nu\sqrt{1-\varrho'^2}}{ 1+\kappa \nu}\pd{\psi}{s}\right)\\
+\pd{}{\nu}\left( \left(\varrho+\nu\sqrt{1-\varrho'^2}\right)(1+\kappa \nu) \pd{\psi}{\nu}\right)
-\frac{m^2(1+\kappa \nu) }{\varrho+\nu\sqrt{1-\varrho'^2}}\psi=0,
\end{multline}
where $\varrho'=d\varrho/ds$ and $\psi$ is considered as a function of $s$ and $\nu$. The interfacial conditions \eqref{transmission} are likewise re-written as 
\begin{equation}\label{fitted transmission}
\psi\ub{i}=\psi\ub{o}, \quad \epsilon\pd{\psi\ub{i}}{\nu}=\pd{\psi\ub{o}}{\nu} \quad \text{at} \quad \nu=0.
\end{equation}

\subsection{Large wavenumber limit}\label{ssec:limit}
Our interest here is in surface-plasmon modes with eigenvalues close to the accumulation point $\epsilon=-1$. We accordingly define the eigenvalue deviation $\tau$ through 
\begin{equation}\label{tau define}
\epsilon=-1-\tau,
\end{equation}
and consider the plasmonic eigenvalue problem in the limit 
\begin{equation}\label{limit}
\tau\to0,
\end{equation}
with $m$ and $f(z)$ fixed. As already discussed in the introduction, in this regime we anticipate modes that oscillate rapidly along the surface, which is why we shall also refer to \eqref{limit} as the large wavenumber limit. Our assumption $m=O(1)$ limits the present study to modes that vary only moderately in the azimuthal direction. 

\section{Surface plasmon rays}\label{sec:rays}
\subsection{Boundary layer and WKBJ ansatz}\label{ssec:BL}
We know, from the structure of Laplace's equation, that the potential must attenuate rapidly away from the interface on a scale which is comparable to the wavelength of the rapid oscillations along the surface. It is not \textit{a priori} evident, however, how that short length scale, say $\delta>0$, relates to $\tau$ except that $\delta(\tau)\to0$ as $\tau\to0$; the sign of $\tau$ is also not obvious. While the solution for a sphere suggests $\delta=O(\tau)$ and $\tau>0$, there are examples of geometries where $\tau<0$ (e.g., even modes for cylinder and sphere dimers) and $\delta\gg\tau$ (e.g., ellipses). We shall here determine the short scale $\delta(\tau)$, along with the sign of $\tau$, as part of the analysis.

To consider the boundary-layer about the particle surface, we write
\begin{equation}\label{bl scaling}
\psi(s,\nu)=\Psi(s,N), \quad N=n/\delta(\tau), 
\end{equation}
and consider the limit $\tau\to0$ with $N$, rather than $\nu$, fixed. Laplace's equation \eqref{fitted laplace} becomes, for $N\ne0$, 
\begin{multline}\label{bl laplace}
\delta^2\pd{}{s}\left(\frac{\varrho+\delta N\sqrt{1-\varrho'^2}}{ 1+\delta \kappa N}\pd{\Psi}{s}\right) +\pd{}{N}\left( \left(\varrho+\delta N\sqrt{1-\varrho'^2}\right)(1+\delta \kappa N) \pd{\Psi}{N}\right)\\ -\frac{\delta^2m^2(1+\delta \kappa N) }{\varrho+\delta N\sqrt{1-\varrho'^2}}\Psi=0
\end{multline}
and the interfacial conditions \eqref{fitted transmission} become 
\begin{equation}\label{bl transmission}
\Psi\ub{i}=\Psi\ub{o} \quad, \quad \pd{\Psi\ub{o}}{N}+\pd{{\Psi}\ub{i}}{N}+\tau\pd{{\Psi}\ub{i}}{N}=0 \quad \text{at} \quad N=0.
\end{equation}
Conditions in the limits $N\to\pm\infty$ are subject to asymptotic matching with a particle-scale outer region, corresponding to the limit $\tau\to0$ with $\nu$ fixed. 

As prompted, the rapid variation in the normal direction must be accompanied by comparably rapid variations along the interface. This, along with the form of \eqref{bl laplace} suggests looking for boundary-layer solutions in the form of a WKBJ ansatz
\begin{equation}\label{ansatz}
\Psi = A(s,N;\tau)\exp\left(\pm\frac{i}{\delta}\int^s k(t)\,dt\right),
\end{equation}
where $A(s,N;\tau)$ is a complex function and $k(s)$ a scaled wavenumber, which without loss of generality is assumed to be positive real.
Substituting  \eqref{ansatz} into \eqref{bl laplace}, we find
\begin{multline}\label{wkb laplace}
\pd{}{N}\left[\left(\varrho+\delta N\sqrt{1-\varrho'^2}\right)\left(1+\delta\kappa N\right)\pd{A}{N}\right]-\frac{k^2\left(\varrho+\delta N\sqrt{1-\varrho'^2}\right)}{1+\delta \kappa N}A \\ 
\pm  i\tau \frac{k \left(\varrho+\delta N\sqrt{1-\varrho'^2}\right)}{1+\delta \kappa N}\pd{A}{s}\pm i\delta \pd{}{s}\left(\frac{\varrho+\delta N\sqrt{1-\varrho'^2}}{1+\delta\kappa N}k A\right)\\
+\delta^2\pd{}{s}\left(\frac{\varrho+\delta N\sqrt{1-\varrho'^2}}{1+\delta \kappa N}\pd{A}{s}\right)-\frac{\delta^2m^2(1+\delta \kappa N)}{\varrho+\delta N\sqrt{1-\varrho'^2}}A=0,
\end{multline}
whereas \eqref{bl transmission} give the conditions
\begin{equation}\label{wkb bcs}
 A\ub{i}=A\ub{o}, \quad \pd{A\ub{o}}{N}+\pd{{A}\ub{i}}{N}+\tau\pd{{A}\ub{i}}{N}=0 \quad \text{at} \quad N=0.
\end{equation}

\subsection{Slowly varying surface plasmons}
Without loss of generality, let the potential in the the boundary layer be $O(1)$. The form of the governing equations \eqref{wkb laplace}--\eqref{wkb bcs} then suggests the asymptotic expansion
\begin{equation}\label{expansions}
A\sim A_0 + \delta A_1 + \delta^2 A_2 + \cdots,
\end{equation}
where at leading $O(1)$ Laplace's equation \eqref{wkb laplace} gives
\begin{equation}\label{A0 eq}
\pd{^2A_0}{N^2}-k^2A_0=0
\end{equation}
and the interfacial conditions \eqref{wkb bcs} give
\begin{equation}\label{A0 bcs}
A_0\ub{i}=A_0\ub{o}, \quad \pd{A_0\ub{i}}{N}+\pd{A_0\ub{o}}{N}=0  \quad \text{at} \quad N=0.
\end{equation}

We see from \eqref{A0 eq}--\eqref{A0 bcs} that $A_0$ either grows or attenuates exponentially as $|N|\to\infty$. Exponential growth contradicts our picture of highly confined surface modes (and clearly leads to a contradiction when matched to the outer region, where the potential satisfies an attenuation condition at large distances from the particle and must remain finite within it). We conclude that $A_0$ attenuates exponentially and hence that the outer potential is exponentially small in $\delta$. This implies, in turn, that the boundary-layer field attenuates at all algebraic orders:
\begin{equation}\label{A decay}
A_l\to0\quad \text{as} \quad |N|\to\infty, \quad l=0,1,2,\ldots
\end{equation}

Note that the leading-order problem \eqref{A0 eq}--\eqref{A decay} is homogeneous; nevertheless, for $s$ fixed, it possesses a two-parameter family of non-trivial solutions in the form 
\begin{equation}\label{A0 sol}
A_0(s,N)=a(s)e^{-k(s)|N|},
\end{equation}
where $a(s)$ is a complex amplitude. Substituting \eqref{A0 sol} into ansatz \eqref{ansatz} shows that the leading-order solution locally resembles a surface plasmon at a flat interface; in fact, the existence of non-trivial solutions to \eqref{A0 eq}--\eqref{A decay} is a consequence of our choice to perturb the permittivity away from $-1$. We next extend the analysis to higher orders in order to determine the slow variation of the wavenumber $k(s)$ and complex amplitude $a(s)$, along with the scaling of $\delta$ and the sign of $\tau$.

\subsection{The short scale $\delta(\tau)$ and the sign of $\tau$}
At $O(\delta)$, Laplace's equation \eqref{wkb laplace} gives 
\begin{equation}\label{A1 eq}
\pd{^2A_1}{N^2}-k^2A_1 = -\frac{1}{\varrho}F_1\{A_0\},
\end{equation}
where we define
\begin{multline}\label{F1 def}
F_1\{A_0\} = \pd{}{N}\left[\left(\sqrt{1-\varrho'^2}+\varrho\kappa\right) N \pd{A_0}{N}\right]-Nk^2\left(\sqrt{1-\varrho'^2}-\varrho\kappa \right)A_0 \\
\pm ik\varrho\pd{A_0}{s}\pm i\pd{}{s}\left(\varrho k A_0\right).
\end{multline}
The form of the interfacial conditions \eqref{wkb bcs} at this order depends on the scaling of $\delta(s)$ and the sign of $\tau$. Let's assume, at first, that $\delta\ll\tau$. Then conditions \eqref{wkb bcs} give 
\begin{equation}\label{A1 bcs wrong 1}
A_1\ub{i}=A_1\ub{o}, \quad \pd{{A}_0\ub{i}}{N}\stackrel{?}{=}0 \quad \text{at} \quad N=0,
\end{equation}
which with \eqref{A0 sol} is a contradiction. If, conversely, we assume that $\delta\gg\tau$, then 
\begin{equation}\label{A1 bcs wrong 2}
A_1\ub{i}=A_1\ub{o}, \quad \pd{A_1\ub{o}}{N}+\pd{A_1\ub{i}}{N}\stackrel{?}{=}0 \quad \text{at} \quad N=0.
\end{equation}
In this case we find an $O(\delta)$ problem, consisting of \eqref{A1 eq}, \eqref{A1 bcs wrong 2} and the decay condition \eqref{A decay}, that is a forced variant of the homogeneous $O(1)$ one. We can therefore only expect solutions to exist under certain ``solvability conditions'' on the forcing terms appearing on the right hand side of \eqref{A1 eq}. By deriving these conditions it can be shown that for finite $\varrho$ and $a\ne0$ the problem for $A_1$ does not posses any solutions. 

We therefore deduce that $\delta(\tau)=O(\tau)$ and without loss of generality  set $\delta=|\tau|$. The transmission conditions \eqref{wkb bcs} give
\begin{equation}\label{A1 bcs}
\pd{A_1\ub{o}}{N}+\pd{A_1\ub{i}}{N}=-\text{sgn}(\tau)\pd{{A}_0\ub{i}}{N}\quad \text{at} \quad N=0
\end{equation}
and we are left to consider the cases of positive and negative $\tau$. The $O(\delta)$ problem is now forced by the right hand sides of both \eqref{A1 eq} and  \eqref{A1 bcs}; for $\tau<0$, it can be shown that solutions exist only for negative $k(s)$, which is again a contradiction. We therefore proceed with
\begin{equation}\label{tau positive}
\delta=\tau>0.
\end{equation} 

\subsection{Relation between wavenumber and local geometry}
With \eqref{tau positive}, the solvability condition for the problem consisting of \eqref{A decay}, \eqref{A1 eq} and \eqref{A1 bcs} can be shown to be 
\begin{equation}\label{k solvability}
k(s) = \frac{\sqrt{1-\varrho'^2(s)}}{\varrho(s)},
\end{equation}
which gives the wavenumber, scaled by the permittivity deviation $\tau$, as a function of the local thickness  of the inclusion. Interestingly, in the case of a sphere, $\varrho(s)=\sin s$, whereby \eqref{k solvability} yields $k(s)\equiv 1$; thus a sphere is a special geometry for which the wavenumber is uniform. The complex amplitude $a(s)$ remains undetermined and so we shall continue to the next order. For this, we require $A_1$. Given the complexity of the expressions, we solve \eqref{A1 eq} and apply the transmission conditions \eqref{A1 bcs} and attenuation \eqref{A decay} using Mathematica \cite{Mathematica}. This yields the solvability condition \eqref{k solvability} --- which can also be derived directly from the governing equations --- along with $A_1$, in which $a(s)$ appears as well as one extra unknown function, say $b(s)$, that does not feature in the subsequent analysis.  

\subsection{Slow variation of amplitude and phase}\label{ssec:slow}
At $O(\tau^2)$, Laplace's equation \eqref{wkb laplace} gives
\begin{equation}\label{A2 eq}
\pd{^2A_2}{N^2}-k^2A_2 = -\frac{1}{\varrho}F_1\{A_1\}-\frac{1}{\varrho}F_2\{A_0\},
\end{equation}
where we define
\begin{multline}\label{F2 def}
F_2\{A_0\} = \pd{}{N}\left(\kappa N^2\sqrt{1-\varrho'^2}\pd{A_0}{N}\right)-k^2\left(\varrho\kappa^2N^2-\kappa\sqrt{1-\varrho'^2}N^2\right)A_0 \\ \pm ik\left(\sqrt{1-\varrho'^2}N-\varrho\kappa N\right)\pd{A_0}{s} 
\pm i\pd{}{s}\left[\left(\sqrt{1-\varrho'^2}-\varrho\kappa\right)NkA_0\right]\\ +\pd{}{s}\left(\varrho\pd{A_0}{s}\right) -\frac{m^2}{\varrho}A_0;
\end{multline}
the transmission conditions \eqref{wkb bcs} give
\begin{equation}\label{A2 bcs}
A_2\ub{i}=A_2\ub{o}, \quad \pd{A_2\ub{o}}{N}+\pd{A_2\ub{i}}{N}=-\pd{{A}_1\ub{i}}{N} \quad \text{at} \quad N=0;
\end{equation}
and $A_2$ also satisfies the attenuation condition \eqref{A decay}. Just as in the previous order, we find a forced variant of the homogeneous $O(1)$ problem. The solvability condition at this order is found to be
\begin{equation}\label{transport}
\pm 2i\frac{da}{ds}+\frac{1-\varrho'^2\pm i\varrho'\sqrt{1-\varrho'^2}}{\varrho\sqrt{1-\varrho'^2}}a=0,
\end{equation}
a differential equation governing $a(s)$. Noting that $a(s)$ is complex, we write
\begin{equation}\label{a decomp}
a=|a|\exp(i\gamma),
\end{equation}
where we later refer to $\gamma$ as the slow phase. 
By splitting \eqref{transport} into real and imaginary parts we obtain differential equations where the amplitude and slow phase appear separately,
\begin{equation}\label{transport separated}
\frac{d}{ds}\left(\varrho|a|^2\right)=0, \quad \frac{d\gamma}{ds}=\pm \frac{1}{2}k,
\end{equation}
 the rate of accumulation of the slow phase being simply proportional to that of the fast phase. Integration of  \eqref{transport separated}, followed by substitution into \eqref{A0 sol}, \eqref{expansions} and \eqref{ansatz}, yields leading-order solutions in the form
\begin{equation}\label{boundary layer}
\psi\sim \text{const.} \times \frac{1}{\sqrt{\varrho(s)}}\exp\left\{\pm i\left(\frac{1}{\tau}+\frac{1}{2}\right)\int^sk(t)\,dt-k(s)|\nu|/\tau \right\},
\end{equation}
where $k(s)$ is provided by \eqref{k solvability}. 

\section{Eigenvalue quantization}\label{sec:quant}
\subsection{General boundary-layer solution}\label{ssec:general}
To generate a general boundary-layer approximation we superimpose the two leading-order solutions in \eqref{boundary layer}:
\begin{equation}\label{bl solution}
\psi \sim \sum_{\pm}\frac{c_{\pm}}{\sqrt{\varrho(s)}}\exp\left[\pm i\left(\frac{1}{\tau}+\frac{1}{2}\right)\theta(s)-k(s)|\nu|/\tau\right].
\end{equation}
Here $c_{\pm}$ are complex constants and we define the phase factor
\begin{equation}\label{theta def}
\theta(s) = \int_{0}^sk(t)\,dt.
\end{equation}
The wavenumber $k(s)$, provided in \eqref{k solvability} in terms of $\varrho(s)$, can also be expressed in terms of the shape function $f(z)$:
\begin{equation}\label{k in zeta}
k(s)=\frac{1}{f(\zeta)\sqrt{1+f'(\zeta)^2}}.
\end{equation}
Since [cf.~\eqref{tips}]
\begin{equation}
k = \kappa_l+o(1) \quad \text{as} \quad z\to-1, \qquad k = \kappa_r+o(1) \quad \text{as} \quad z\to1,
\end{equation}
and since $f>0$ for $|z|<1$, $\theta(s)$ is well defined by \eqref{theta def} and is a monotonically increasing function of $s$; using \eqref{arc length} we can write $\theta(s)$ as 
\begin{equation}\label{theta def}
\theta(s) = \int_{-1}^{\zeta(s)}\frac{dz}{f(z)}.
\end{equation}
We shall later find the geometric parameter
\begin{equation}\label{big theta}
\Theta=\int_{-1}^{1}\frac{dz}{f(z)}
\end{equation}
to play an important role in the selection of discrete eigenvalues; note that $\Theta$ equals the product of the inclusion length and the reciprocal average of its cross-sectional radius along its symmetry axis. 

We interpret the leading-order boundary-layer solution \eqref{bl solution} as being formed of left- and right-going surface-plasmon rays that traverse the meridian arcs (note that, since $m=O(1)$, the azimuthal wavenumber is small relative to the longitudinal one). At this stage, the coefficients $c^{\pm}$ and the eigenvalue deviation $\tau$ remain undetermined. Clearly \eqref{bl solution} is singular at the poles, where $\varrho=0$. At these singularities, the surface plasmon rays focus and the boundary-layer scaling \eqref{bl scaling} and the WKBJ ansatz \eqref{ansatz} break down. We next show that \eqref{bl solution} can be matched with inner regions near to the poles, and that this quantizes $\tau$ and determines the coefficients $c^{\pm}$ (up to a common  multiplicative constant).

\subsection{Matching with pole regions}\label{ssec:poles}
Consider first the inner region at $O(\tau)$ distances from the pole at $z=-1$. Noting that $\varrho\sim s$ as $s\to0$, for $s=O(\tau)$ the boundary-layer solution \eqref{bl solution} becomes $O(\tau^{-1/2})$ large. We  accordingly expand the potential as
\begin{equation}\label{inner expansion}
\psi(s,\nu;\tau)\sim \tau^{-{1/2}}\Psi_l(S,N)
\end{equation}
and consider the inner limit where the stretched co-ordinates $S=s/\tau$ and $N$ are fixed. The leading inner potential $\Psi_l$ satisfies Laplace's equation [cf.~\eqref{fitted laplace}]
\begin{equation}\label{inner laplace}
\frac{1}{S}\pd{}{S}\left(S\pd{\Psi_l}{S}\right)+\pd{^2\Psi_l}{N^2}-\frac{m^2}{S^2}\Psi_l= 0, \quad N\ne0;
\end{equation}
along with the transmission conditions [cf.~\eqref{transmission}]
\begin{equation}\label{inner transmission}
\Psi_l\ub{o}= \Psi_l\ub{i}, \quad \pd{\Psi_l\ub{o}}{N}+\pd{\Psi_l\ub{i}}{N}= 0 \quad \text{at} \quad N=0,
\end{equation}
which do not involve the eigenvalue deviation $\tau$; the attenuation conditions
\begin{equation}\label{inner decay}
\Psi_l\to0 \quad \text{as} \quad |N|\to\infty
\end{equation}
as well as the condition of asymptotic matching with the boundary-layer solutions in the limit $S\to\infty$. Regarding the latter, we note that the inner limit of the leading $O(1)$ boundary-layer solution \eqref{bl solution},
\begin{equation}\label{inner of outer}
\sim (\tau S)^{-1/2}e^{-\kappa_l |N|}\sum_{\pm}c_{\pm}e^{\pm i\kappa_lS},
\end{equation}
should match with the corresponding boundary-layer limit of the inner expansion. 

Separable solutions that satisfy \eqref{inner laplace}--\eqref{inner decay} and are regular at $S=0$ are proportional to $J_m(\beta S)\exp(-\beta|N|)$, where $\beta$ is an arbitrary separation constant. Since 
\begin{equation}\label{bessel}
J_m(\kappa_lS)\sim \frac{1}{\sqrt{2\pi \kappa_l S}}\sum_{\pm}e^{\pm i\left(\kappa_l S-\alpha_m\right)}
\quad \text{as} \quad S\to\infty, 
\end{equation}
where
\begin{equation}\label{alpha}
\alpha_m=\frac{m\pi}{2}+\frac{\pi}{4},
\end{equation}
the need to match with \eqref{inner of outer} implies that the requisite inner solution is simply 
\begin{equation}\label{pole left}
\Psi_l(S,N) = c_l J_m(\kappa_l S)e^{-\kappa_l|N|},
\end{equation}
where the pre-factor $c_l$ is linked to the boundary-layer coefficients through
\begin{equation}\label{cl match}
c_{\pm}=c_l\frac{e^{\mp i\alpha_m}}{\sqrt{2\pi\kappa_l}}.
\end{equation}
The above relations determine the boundary-layer solution in terms of a single multiplicative constant, say $c_l$ (and the eigenvalue deviation $\tau$). 

We next consider the pole region near $z=1$. The relevant inner limit is with $\tilde{S}=(s_{z=1}-s)/\tau$ and $N$ fixed. The problem formulation is similar to the first inner region and hence
\begin{equation}\label{other inner}
\psi\sim \tau^{-1/2}\Psi_r(\tilde{S},N), \quad \Psi_r(\tilde{S},N)= c_rJ_m(\kappa_r\tilde{S})\exp(-\kappa_r|N|),
\end{equation}
where $c_r$ is an unknown pre-factor. Similar to before, the inner limit of the boundary-layer solution \eqref{bl solution}, 
\begin{equation}\label{inner of outer right}
\sim (\tau \tilde{S})^{-1/2}e^{-\kappa_r |N|}\sum_{\pm}c_{\pm}e^{\pm i\left(\frac{1}{\tau}+\frac{1}{2}\right)\Theta\mp i\kappa_r\tilde{S}},
\end{equation}
where we used $\theta(s)\sim \Theta- \kappa_r(s_{z=1}-s)$ as $z\to1$, should match with the boundary-layer limit of \eqref{other inner}, obtained using \eqref{bessel} with $\kappa_l S$ replaced by $\kappa_r\tilde{S}$. This matching condition links $c_r$ to the boundary-layer coefficients:
\begin{equation}\label{cr match}
\frac{c_r e^{\pm i\alpha_m}}{\sqrt{2\pi \kappa_r}}=c_{\pm}e^{\pm i\left(\frac{1}{\tau}+\frac{1}{2}\right)\Theta}.
\end{equation}

\subsection{Global solvability}
At this stage we have determined the global form of the solution, which is composed of the boundary-layer solution \eqref{bl solution}, characterized by the coefficients $c_{\pm}$, and the solutions in the two inner pole regions, which are respectively characterized by coefficients $c_l$ and $c_r$. These four coefficients are related through the four matching conditions provided in \eqref{cl match} and \eqref{cr match}, which can be combined to give the matrix equation 
\begin{equation}\label{matrix}
\left(\begin{array}{cccc}
\sqrt{2\pi\kappa_l} & 0 & -{e^{-i\alpha_m}} & 0 \\
0 & \sqrt{2\pi\kappa_l} & -e^{i\alpha_m} & 0 \\
\sqrt{2\pi\kappa_r}e^{i\Theta\left(\frac{1}{\tau}+\frac{1}{2}\right)} & 0 & 0 & -e^{i\alpha_m} \\
0 & \sqrt{2\pi\kappa_r}e^{-i\Theta\left(\frac{1}{\tau}+\frac{1}{2}\right)} & 0 & -e^{-i\alpha_m} \end{array}\right)
\left(\begin{array}{c} c_+ \\ c_- \\ c_l \\ c_r \end{array}\right)= \left(\begin{array}{c} 0 \\ 0 \\ 0 \\ 0 \end{array}\right).
\end{equation}
For a non-trivial solution the determinant of the matrix must vanish, which yields the solvability condition [cf.~\eqref{alpha}]
\begin{equation}\label{quant solvability}
\sin\left(\frac{\pi}{2}-\Theta\left(\frac{1}{\tau}+\frac{1}{2}\right)+m\pi\right)=0.
\end{equation}
This condition, in turn, implies the asymptotic quantization rule
\begin{equation}\label{quant}
\frac{1}{\tau}\sim \frac{\pi n }{\Theta}+\frac{1}{2}\left(\frac{\pi}{\Theta}-1\right)+o(1) \quad \text{for} \quad  1\ll  n \in \mathbb{N}. 
\end{equation}
For each asymptotic eigenvalue, solving \eqref{matrix} yields a leading-order approximation for the corresponding eigen-potentials. 

\begin{table}[t!]
\setlength\tabcolsep{10pt}
\begin{center}\scalebox{0.9}{
\begin{tabular}{|c|c|c|c|c|c|c|c|}
\hline
& \multicolumn{4}{|c|}{$h=0.7$} & \multicolumn{3}{|c|}{$h=0.25$}\\
\hline
n & $n\gg1$ & $m=0$ &  $m=1$  & $m=2$ & 	      $n\gg1$ & $m=0$ & $m=1$ \\
\hline
     1  &		2.818            &       3.096  & 1.646  &   1.227           & -7 & 12.261 & 1.163 \\
     2  &		1.8                &       1.885   & 1.558 &  1.295    & 9  &  5.718 & 1.281\\
     3  &		1.513            &       1.541   & 1.437  &  1.294   & 3.667 & 3.745 & 1.365\\
     4  &		1.377    	    &       1.389    & 1.347  &  1.269   & 2.6  & 2.843 & 1.415\\
     5  &		1.299           &       1.304    &  1.284  & 1.238    & 2.143 & 2.344 & 1.437\\
     6  &		1.247           &       1.250    &   1.239 & 1.211    & 1.889  & 2.036 & 1.439\\
     7  &		1.211          &       1.212    &  1.205    &  1.187    &  1.727 & 1.833 & 1.428\\
     8  &		1.183          &       1.185    &   1.180  &   1.168     &  1.615  &  1.691 & 1.410\\
     9  &		1.163          &       1.163    &   1.160  &   1.151     &   1.533  &  1.588 & 1.390\\
    10  &		1.146          &     1.146     &   1.144  &     1.138   &  1.471  &  1.510  & 1.367\\
    11  &		1.132          &     1.133     &   1.131 &    1.126    &  1.421  &   1.450  & 1.344\\
    12  &		1.121          &     1.121     &   1.120 &    1.116   &  1.381  &   1.403  & 1.324\\
   13  &		1.112          &     1.112     &   1.111 &    1.108   &  1.348   &   1.365  & 1.304\\
   14  &		1.104          &     1.104     &   1.103 &    1.101     &   1.32    &  1.333  & 1.286\\
   15  &		1.097          &     1.097     &   1.096 &    1.094      &   1.296   &  1.306  & 1.269\\
     16  &		1.091          &     1.091     &   1.090 &    1.088     &    1.276   &  1.284 & 1.254\\
     17  &		1.085          &     1.085     &   1.085 &    1.083    &    1.258    &  1.265 & 1.240\\   
     18  &		1.080          &     1.080     &   1.080 &    1.079    &     1.242  &   1.248 & 1.228\\  
     19  &		1.076          &     1.076     &   1.076 &    1.075    &     1.229   &  1.233  & 1.216\\ 
     20  &		1.072          &     1.072     &   1.072 &    1.071    &    1.216   & 1.220  & 1.206 \\
     21  &		1.069          &     1.069     &   1.069 &    1.068    &    1.205   & 1.208  & 1.196 \\
     22  &		1.066          &     1.066     &   1.065 &    1.065    &    1.195   & 1.198  & 1.188 \\
     23  &		1.063          &     1.063     &   1.063 &    1.062    &    1.186   & 1.188  & 1.180 \\
\hline
\end{tabular}}
\end{center}
\caption{Eigenvalues $-\epsilon$ for a prolate spheroid of thickness to length ratio $h$. The asymptotic values for $n\gg1$, provided by \eqref{quant} with $\Theta=\pi/h$, are compared with exact values for $m=0,1$ and $2$, for two values of $h$.}
\label{table}
\end{table}%
\section{Examples: Spheres, spheroids and asymmetric shapes}\label{sec:examples}
It is illuminating to apply our asymptotic theory to the case of a sphere, for which the eigenvalues and eigenfunctions are known in closed form. The thickness profile for a sphere is $f(z)=\sqrt{1-z^2}$, whereby the integral \eqref{big theta} gives $\Theta=\pi$. The leading term in the quantization rule \eqref{quant} thus returns the exact eigenvalues \eqref{ev sphere}, with the correction term accordingly vanishing. The corresponding mode for given $m$ and $n$ is determined by solving \eqref{matrix}; let $c_l=1$, then $c_r=(-1)^{m+n}$ and $c_{\pm}=(2\pi)^{-1/2}e^{\mp i \alpha_m}$. The corresponding boundary-layer solution \eqref{bl solution} and pole solutions \eqref{pole left} and \eqref{other inner} together constitute a leading-order approximation for the eigenfunctions of a sphere, which generalize the asymptotic structure discussed in the introduction, from $m=0$ to arbitrary $m=O(1)$. 

Consider next a prolate spheroid, in which case $f(z)=h\sqrt{1-z^2}$, where $h$ is the ratio of the minor and major axes. From \eqref{big theta} we find $\Theta=\pi/h$, with the asymptotic eigenvalues and eigenfunctions following from \eqref{quant} and \eqref{matrix} similarly to before; note that in this case the second term in \eqref{quant} does not vanish.
In table \ref{table} we compare the eigenvalues predicted by \eqref{quant} with the exact eigenvalues, which are known in terms of associated Legendre functions of the first and second kind (see, e.g., \cite{Klimov:14}). The rate of convergence of the exact values to the asymptotic ones with increasing $n$ is seen to be slower for smaller values of $h$. In fact, it is evident from \eqref{quant} that our asymptotic analysis fails in the slender-body limit $h\ll1$ as the two terms in \eqref{quant} become comparable; this non-commutativity of limits is to be expected, given that the longitudinal  modes of elongated particles tend to $-\infty$, rather than $-1$, in the slender-body limit. The convergence of the asymptotic values to the exact ones is also seen to decelerate with increasing $m$, especially for small $h$. As $m$ is increased and $h$ is made smaller, the surface-plasmon rays deviate from the meridian arcs and the modes gradually approach the 2D modes of a circular cylinder, which all share the eigenvalue $\epsilon=-1$. The large $|m|$ and slender-body limits are further discussed in \S\ref{sec:concluding}. 

Of course, it is just as easy to apply the theory to other permissible shapes for which the exact solution is unavailable. To give one example, for an asymmetric inclusion whose thickness profile is $f(z)=h(p+z)\sqrt{1-z^2}$, where $p>1$, we find from \eqref{big theta} that $\Theta=({\pi/h})/{\sqrt{p^2-1}}$. 

\section{Concluding remarks}\label{sec:concluding}
Plasmonic modes of smooth axisymmetric inclusions, with eigenvalues close to accumulation and moderate azimuthal number $m$, can be interpreted as being composed of surface-plasmon rays traversing the meridian arcs of the boundary. In this description, the modes are confined to a narrow boundary layer about the interface, where the eigen-potentials are locally   surface plasmons of a flat interface. The wavenumber and amplitude slowly vary along the rays, as a function of the local thickness of the inclusion; in addition to the fast phase oscillations, there is also a slow phase accumulation associated with the geometry and the azimuthal dependence. At the poles the boundary-layer fields diverge and are seen to match with inner  regions where the surface plasmons focus and are amplified. The WKBJ analysis leading to the surface-plasmon ray description is somewhat unconventional in that three (rather than two) consecutive orders need to be considered just to determine the leading-order solution. This is because the local wavenumber is not known \textit{a priori}, as in traditional geometric optics, for example, but rather is governed by the geometry and must be determined as part of the analysis. 

By matching the right- and left-going surface-plasmon solutions in the boundary layer with the two inner pole regions, the original eigenvalue problem is reduced to a matrix equation governing a set of four scalar coefficients. Solvability of the reduced problem yields the two-term quantization rule \eqref{quant} for the eigenvalue deviation. The leading order term simply states that a whole number $n$ of plasmon half-wavelengths fit along the meridian arc connecting the inclusion poles. In the special case of a sphere, the wavenumber is uniform and equal to $1/\tau$, so that this condition reads $n\times (\pi/\tau) = \pi$, which remarkably yields the exact eigenvalues \eqref{ev sphere}. For more general shapes the wavenumber varies over the surface and the effective plasmon path is proportional to the geometric parameter $\Theta$. The correction term in \eqref{quant}, which is also given in terms of $\Theta$, is due to the slow phase accumulation along the rays and at the two poles. We note that in the physics literature a different quantization rule has been proposed, where the wavenumber is not determined by the local geometry but rather from the dispersion relation of an electromagnetic surface wave (surface-plasmon polariton) propagating along a flat surface \cite{Fang:14,Liu:15}; that description is only relevant to high-order modes of large inclusions, i.e., outside the quasi-static regime. 

Curiously, the two leading terms in the quantization rule \eqref{quant} are independent of the azimuthal number $m$, even though the latter affects the slow accumulation of phase both along the rays and at the poles. In the case of a sphere, we know from the exact solution \eqref{ev sphere} that this remains true to all asymptotic orders. The exact solution in the case of a spheroid demonstrates, however, that this is not the case in general. While in the present paper the asymptotic analysis was carried out with $m$ fixed, we expect that a similar approach could be applied to study the large-$|m|$ modes. In the latter case, the plasmon rays deviate from the meridian arcs and, instead of focusing at the poles, they bounce back and forth between circular caustics, defined by the intersection of the boundary with certain constant-$z$ planes. We note that without knowledge of the large-$|m|$ eigenvalue asymptotics it does not seem possible to compare (the leading term of) the quantization rule \eqref{quant} with the Weyl's law proved in \cite{Miyanishi:18}. 

The thickness profile $f(z)$ was also held fixed in our asymptotic analysis. Accordingly, our asymptotic approximations are not useful for rapidly varying, or very slender, shapes, unless exceedingly high mode numbers are considered; examples of such inclusions include finite cylinders truncated over a short scale, like the nano rods commonly used in plasmonic applications \cite{Chen:13}, and high-aspect-ratio needle-like inclusions, like the small-$h$ prolate spheroid discussed in \S\ref{sec:examples}. Singular perturbation techniques \cite{Hinch:91}, especially slender-body theory, could be used to find asymptotic solutions in the latter limits, for both low-order and high-order modes. 

We focused in this paper on smooth axisymmetric shapes, for which the axial symmetry allows decoupling the azimuthal dependence and the formation of ray singularities is fairly evident. Whereas deriving the surface-plasmon-ray equations for an arbitrary 3D interface would probably not be much more difficult than the axisymmetric analysis in \S\ref{sec:rays}, inferring the global structure of the solution for general shapes would be intractable in most cases. For smooth planar inclusions in 2D, however, it is clear that there is just a single ray, that traverses the boundary with no singularity forming. It can be shown that in that case the boundary-layer analysis leaves the eigenvalues undetermined at all algebraic orders. Indeed, for smooth planar inclusions the rate of convergence to the accumulation point is exponential \cite{Ando:16exp}. In the ray-theory picture, it seems necessary to account for the weak self-interaction of the exponential tail of the surface-plasmon ray with the boundary. The techniques of exponential asymptotics \cite{Vanden:02} may be useful in addressing the 2D high-mode-number limit.

We lastly note that with increasing mode number the plasmon wavelength ultimately becomes commensurate with the scale of surface roughness or the scale at which local electromagnetic theory breaks down. In most plasmonic excitation problems, where only low-order modes significantly contribute to the optical response, these deviations from ideality are immaterial. In scenarios where high-mode-number are important, however, it may be important to account for roughness and nonlocal corrections. A generalized plasmonic eigenvalue problem that approximately accounts for nonlocality can be formulated based on the hydrodynamic Drude model \cite{Luo:13,Raza:15}. A perturbation analysis of this problem in the limit where the Fermi screening length is small compared to the inclusion size confirms that, for moderate mode numbers, the effect of nonlocality is weak but becomes more pronounced as the mode number increases \cite{Schnitzer:16b}. It is clear that for high mode numbers, where the plasmon wavelength is comparable with the screening length, nonlocality is no longer a perturbative effect and that the accumulation of eigenvalues predicted in the local case is ultimately regularized. It would therefore be interesting to analyze the nonlocal eigenvalue problem in the double limit of high-mode number and small Fermi-screening length.

\textbf{Acknowledgements.} The author acknowledges funding from EPSRC through New Investigator Award EP/R041458/1.

\appendix
\section{Boundary-fitted co-ordinates}
Our goal here is to derive the form of Laplace's equation in terms of the boundary-fitted co-ordinates $(s,\nu,\phi)$ defined in \S\ref{sec:formulation}. Writing the arc length as 
\begin{equation}\label{arc length}
s=\int_{-1}^{\zeta}\sqrt{1+f'(t)^2}\,dt 
\end{equation}
and using $f(\zeta)=\varrho(s)$, we find 
\begin{equation}\label{f rho relations}
f'(\zeta) = \frac{d\varrho}{ds}\sqrt{1+f'(\zeta)^2}, \qquad f'(\zeta)^2= \left(1- \varrho'(s)^2\right)^{-1}\varrho'(s)^2.
\end{equation}
We next parameterise the boundary as
\begin{equation}\label{surface position}
\bx_S(s,\phi)=f(\zeta)\be_{\rho}(\phi)+\zeta\be_{z},
\end{equation}
where ($\be_{\rho},\be_z,\be_{\phi}$) are the unit vectors associated with the cylindrical co-ordinates $(\rho,z,\phi)$. We also write the normal unit vector, pointing into the background domain, as 
\begin{equation}\label{normal}
\bn(s,\phi) 
= \left(\be_{\rho}(\phi)-f'(\zeta)\be_z\right)\left(1+f'(\zeta)^2\right)^{-1/2}. 
\end{equation}
In some finite neighbourhood of the boundary, the position of any point $\bx= z\be_z+r\be_r(\phi)$ can be written as
\begin{equation}\label{position}
\bx  = \bx_S(s,\phi)+\nu \bn(s,\phi),
\end{equation}
which gives the transformation between the cylindrical and boundary-fitted co-ordinates. In particular, our interest is in the scale factors 
\begin{equation}\label{scale1}
h_s = \left|\pd{\bx}{s}\right|, \quad h_{\nu} = \left|\pd{\bx}{\nu}\right|, \quad h_{\phi} = \left|\pd{\bx}{\phi}\right|,
\end{equation}
where $\bx$ here is treated as a function of the boundary-fitted co-ordinates. 
Substituting \eqref{position} into \eqref{scale1} gives $h_{\nu}=1$ and 
\begin{equation}\label{scale2}
h_s = \left|\pd{\bx_S}{s}+\nu\pd{\bn}{s}\right|, \quad h_{\phi} = \left|\pd{\bx_S}{\phi}+\nu\pd{\bn}{\phi}\right|.
\end{equation}
Noting that
\begin{equation}\label{curve equations}
\pd{\bn}{s} = \kappa \bt,
\end{equation}
where $\bt=\partial{\bx_S}/\partial{s}$ is a tangent unit vector and $\kappa(s)$ is the signed curvature of a meridian arc 
such that $\kappa(s)\gtrless 0$ wherever $f''(\zeta)\lessgtr 0$, we find
\begin{equation}\label{h s}
h_s(s,\nu)=1+\kappa(s) \nu.
\end{equation}
To similarly simplify $h_{\phi}$ we use \eqref{surface position}, \eqref{normal} and \eqref{scale2} to find 
\begin{equation}\label{h phi}
h_{\phi}(s,\nu) = \left|\varrho(s)\pd{\be_{\rho}}{\phi}+\nu\pd{\bn}{\phi}\right|=\left|\varrho(s)+\nu\left(1+f'(\zeta)^2\right)^{-1/2}\right|,
\end{equation}
which with \eqref{f rho relations} reduces to 
\begin{equation}\label{h phi}
h_{\phi}(s,\nu) =\varrho(s)+\nu\left({1-\varrho'(s)^2}\right)^{1/2}
\end{equation}
for $\nu$ not too large. Using the above expressions for the scale factors, the Laplacian operator can be written as
\begin{multline}\label{fitted laplacian}
\nabla^2\varphi = \frac{1}{\left(\varrho+\nu\sqrt{1-\varrho'^2}\right) (1+\kappa \nu)}\left[\pd{}{s}\left(\frac{\varrho+\nu\sqrt{1-\varrho'^2}}{ 1+\kappa \nu}\pd{\varphi}{s}\right)\right. \\ \left. +\pd{}{\nu}\left( (\varrho+\nu\sqrt{1-\varrho'^2})(1+\kappa \nu) \pd{\varphi}{\nu}\right)+\frac{ 1+\kappa \nu }{\varrho+\nu\sqrt{1-\varrho'^2}}\pd{^2\varphi}{\phi^2}\right].
\end{multline}

\bibliographystyle{unsrt}
\bibliography{refs}
\end{document}